\documentclass[fleqn,usenatbib]{mnras}

\usepackage[T1]{fontenc}
\usepackage{graphicx}	
\usepackage{amsmath}	
\usepackage{amssymb}	
\usepackage{booktabs}
\usepackage{diagbox}
\usepackage{multirow}
\usepackage{color}
\usepackage{soul}
\usepackage{float}
\usepackage{graphicx}	
\usepackage{amsmath}
\usepackage{epstopdf}
\usepackage{array}
\usepackage{lineno}
\usepackage{soul}
\soulregister{\cite}7
\soulregister{\citep}7 
\soulregister{\citet}7
\soulregister{\ref}7 
\soulregister{\pageref}7 

\title[time-integrated polarizations]{Revisiting the time-integrated polarizations of gamma-ray burst prompt phase}

\author[Sui and Lan]{
Li-Qiang Sui$^1$ and
Mi-Xiang Lan$^{1}$
\\
\\
$^{1}$ Center for Theoretical Physics and College of Physics, Jilin University, Changchun, 130012, China; lanmixiang@jlu.edu.cn}

\date{Accepted XXX. Received YYY; in original form ZZZ}
\pubyear{2023}
\let\oldequation\equation
\let\oldendequation\endequation
\renewenvironment{equation}
{\linenomathNonumbers\oldequation}
{\oldendequation\endlinenomath}

\begin{document}

\label{firstpage}
\pagerange{\pageref{firstpage}--\pageref{lastpage}}
\maketitle
\begin{abstract}
In the former studies, the time evolution information is missed in deducing the time-integrated polarizations of gamma-ray burst (GRB) prompt emission. Here, it is considered and the time-integrated polarizations is investigated through the summation of the time-resolved ones. The statistical properties of the distribution of the time-integrated polarization degree ($\Pi$) can be read from the $q-\Pi$ curve, where $q\equiv\theta_V/\theta_j$. $\theta_V$ and $\theta_j$ are the observational and jet half-opening angles, respectively. Hence, only the $q-\Pi$ curves are studied. In addition to a toroidal magnetic field in the radiation region, an aligned field is also discussed. We found the predicted time-integrated PD is around $(40-50)\%$ for High-energy Polarimetry Detector (HPD) on board POLAR-2 and is roughly $(30-40)\%$ for its Low-energy Polarimetry Detector (LPD). Therefore, $\Pi$ value detected by the HPD will be larger than that of the LPD in statistics and the result of the former estimations will underestimate the value of $\Pi$ in an ordered field. There are mainly two types of the $q-\Pi$ curve profiles, corresponding to two ordered magnetic field configurations.
\end{abstract}



\section{Introduction} \label{sec:intro}
Gamma-ray bursts (GRBs) are the extensive $\gamma$-ray radiation at the cosmological distance, with extremely high isotropic-equivalent luminosity of $L_{iso}\sim 10^{50}-10^{53}\ erg/s$ \citep{Kumar20151}. After more than two decades intensive studies, their emission mechanism \citep{Rees1994,Paczynski_and_Xu1994,Giannios2008,ZY2011,Granot2016,Thompson1994,Eichler1999,Eichler_2003,Lundman2018}, jet structure \citep{Rossi2004,Gill2020}, and magnetic field configuration \citep{Sari1999,Granot2003a,Granot2003b,Toma2009,Lan2019} (MFC) in the emitting regions remain mysterious. The light curves of GRB prompt emission usually show abundant diversity. Different from the varieties of GRB prompt light curves, their spectra can be described by the Band function \citep{Band1993}. However, the origin of the Band function is also a mystery. The Band function is an empirical function, consisting of low- and high-energy power laws, smoothly connected around the peak energy $E_{p}$ of the $\nu F_{\nu}$ spectrum \citep{Band1993}, which can not be interpreted by the fast-cooling synchrotron radiation in a constant magnetic field. Then the synchrotron origin of the Band spectrum in a decaying magnetic field was proposed \citep{PZ2006,Derishev2007,Zhao2014,Uhm_Z.Lucas2014}.

To explain the observations of the GRB prompt emission, three popular models were proposed, i.e., the internal shock (IS) model \citep{Rees1994,Paczynski_and_Xu1994}, the magnetic reconnection (MR) model \citep{Giannios2008,ZY2011,Granot2016}, and the photosphere model \citep{Thompson1994,Eichler1999,Eichler_2003,Lundman_Pe'er2013}. The radiation mechanism of the first two models are the synchrotron emission, while it might be synchrotron emission in the low-energy X-ray band and be the inverse Compton scattering in the high-energy $\gamma-$ray band for the photosphere model \citep{Lundman2018}. The emission model of GRB prompt phase proposed by \cite{Uhm_Zhang2015, Uhm_Zhang2016} and \cite{Uhm_2018} was used here. It can mimic the emission from both the MR process and the IS process.

Up till now, most of the polarization observations in GRB prompt phase were given as the time- and energy-integrated ones, and the time-resolved polarization results are relatively rare \citep{Yonetoku2011,Yonetoku2012,Zhang2019,Kole2020,Sharma2019,Chattopadhyay2022}. Theoretically, time-resolved and time-integrated polarizations of GRB prompt emission were considered widely in the literature \citep{Granot2003b,Lyutikov2003,Nakar2003,Lazzati2004,GT2005,Toma2009,Lundman2014,Gill2020,LD2020,Cheng2020,GG2021,LWD2021,Lan2021,GL2023}. In these above mentioned time-integrated models, the time-integrated Stokes parameters were directly constructed from the observed energy spectrum. So the time-evolving information of the polarization was not included. Recently, \cite{GG2024} considered the time-integrated polarizations in GRB prompt phase via the summation of the time-resolved ones with evolving Polarization angle (PA).

A rotating PA during the burst, which was already observed in 4 GRBs \citep{Yonetoku2011,Zhang2019,Sharma2019,Kole2020,Chattopadhyay2022}, can reduce the final time-integrated polarization degree (PD), as in the case of GRB 170114A \citep{Zhang2019,Burgess2019,Kole2020}. And the order degree of the magnetic field also decays during the main radiation episode, which would be the case for the MR process. In addition, the energy spectrum would evolve during GRB prompt phase, which could affect the time-integrated PD of the jet emission \citep{GL2023}. Therefore, time-resolved polarization information will be important for the final time-integrated ones and should be considered. 

Here, we consider the time- and energy-resolved polarization using the polarization model proposed in \cite{LD2020} with a top-hat jet structure to investigate the time- and energy-integrated PD of GRB prompt phase. POLAR-2 is a forth coming mission dedicated to detect the GRB polarizations in both X-ray and $\gamma-$ray band \citep{POLAR2}. The detection energy band of the two polarization detectors on board POLAR-2, the High-energy Polarimetry Detector (HPD) and Low-energy Polarimetry Detector (LPD), are 30-800 keV and 2-30 keV, respectively. We apply our model to predict the detection results of the HPD and LPD on board POLAR-2. The dependence of the results on the parameters are also investigated. This paper is arranged as follows. In Section 2, the polarization model is reviewed. The time-integrated PDs at energy band of LPD (2-30 keV) and HPD (30-800 keV) are presented in Section 3. Finally, we give our conclusion and discussion in Section 4.

\section{The model}\label{sec:The model}

\begin{figure}
  \centering
  \includegraphics[scale=0.7]{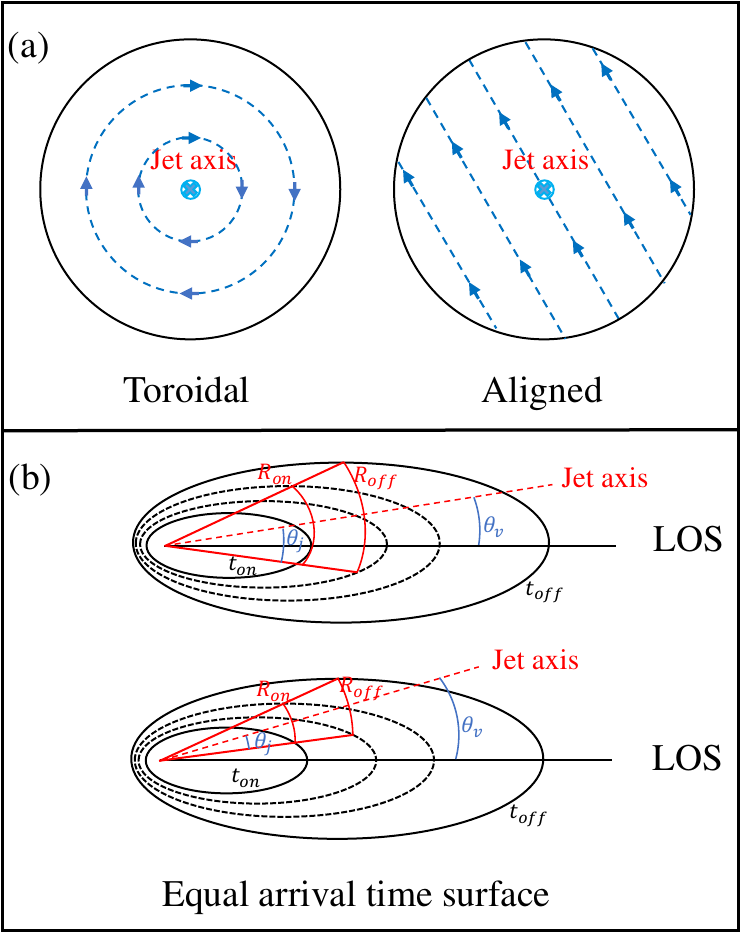}
  \caption{(a) Schematic diagrams for the two kinds of the ordered magnetic field in the jet surface. The blue arrows show the field directions. And the cross represents the jet axis. (b) Schematic cross-sections of the EATSes at different times for on-axis (top) and off-axis (bottom) observations. The radiation starts from $t_{on}$ and ceases at $t_{off}$.}
\label{Fig1}
\end{figure}

The emission model used here was proposed by \cite{Uhm_Zhang2015, Uhm_Zhang2016} and \cite{Uhm_2018}. A thin, radially expanding relativistic jet shell continually emits photons from all positions in the shell. It begins to radiate at radius $r_{on}$ and stops at radius $r_{off}$. In the comoving frame of the shell, it is assumed that the radiation power has an isotropic angular distribution. The jet is assumed to have a top-hat structure with a half-opening angle of $\theta_j$. Under the framework of the MR model, as the highly magnetized jet shell moves radially outward, because of the instabilities, the field with modulations would reconnect and release free magnetic energy to accelerate the jet and the electrons in it. There are two kinds of ordered magnetic field discussed in the literature of MR: aligned and toroidal \citep{Spruit2001}. The field lines of the aligned field are the latitude circles on the jet surface \citep{Spruit2001,Lan2019}, while they are concentric circles around the jet axis for the toroidal field \citep{Spruit2001}. The schematic diagrams of the two field configurations, the EATS, and the relevant angles are shown in Figure \ref{Fig1}. 

The Blandford-Znajek (BZ) mechanism would work for a black hole central engine and the field configuration in the jet at large radii would be toroidal \citep{Blandford1977}, while for a magnetar a striped wind is ejected via magnetic dipole radiation and MFC is aligned in the ejecta \citep{Spruit2001}. An aligned field is unstable to the tearing instability, while a toroidal field is highly unstable to the Kink instability. The turbulence would develop and the fields with different orientations will reconnect. Then the jet will be accelerated by the released free magnetic energy of the MR process. The bulk Lorentz factor would vary in the form of power laws with respect to the radius \citep{Drenkhahn2002}. 
\begin{equation}
\Gamma(r)=\Gamma_0(r/r_0)^s.
\end{equation}
It was predicted that bulk acceleration patterns will vary with $s=0$ for the toroidal field and with $s=1/3$ for the aligned field under the MR model \citep{Drenkhahn2002}. And the magnetic field strength in the co-moving frame of the radiating region will decrease with radius due to the adiabatic expansion of the jet shell and the MR process.
\begin{equation}
B'(r)=B'_0(r/r_0)^{-b}.
\end{equation}

The emission model discussed above is not only suitable to the MR process, but also can mimic the emission of the internal shock (IS) model. In the scenario of the IS, the electrons are accelerated by the ISs and injected into the shell continuously until the shock crossed the shell. The magnetization would be lower than $10^{-3}$ in order for the shocks to accelerate particles and the magnetic field would be dominated by the shock-generated small-scale random field \citep{SS2011}. The decay of the ordered magnetic field for the MR process should be faster than $b=1$ because besides the adiabatic expansion of the field with the jet shell it will also reconnect to deplete the free magnetic energy. So the field would be mixed with both ordered component and random component in the MR region. Another difference between the MR model and the IS model is that the bulk velocity of the shell will be constant for the IS (i.e., $s=0$), while it can increase with radius for the MR model.

The polarization of the above emission model was predicted by \citet{LD2020}, which was used here for the time-resolved polarization calculation. The expressions of the time- and energy-resolved Stokes parameters are slightly different from that in \cite{LD2020},  we type these formula in the Appendix for reference. Then the energy-integrated Stokes parameters are derived with the following formula.
\begin{equation}
F=\int_{\nu_1}^{\nu_2}F_{\nu} d\nu
\end{equation}
\begin{equation}
Q=\int_{\nu_1}^{\nu_2}Q_{\nu} d\nu
\end{equation}
\begin{equation}
U=\int_{\nu_1}^{\nu_2}U_{\nu} d\nu
\end{equation}
where $F_{\nu}$, $Q_{\nu}$, and $U_{\nu}$ are the time- and energy-resolved Stokes parameters. $\nu$ represent the observational frequency. $h\nu_1$ and $h\nu_2$ are the lower- and upper-limits of the corresponding detector, respectively. $h$ is the Planck constant. Here, $h\nu_1=30$ keV and $h\nu_2=800$ keV for the HPD, and $h\nu_1=2$ keV and $h\nu_2=30$ keV for the LPD.

Since the flux and polarized flux are very tiny without $T_{90}$, we neglect these contributions to the time-integrated polarization. Therefore, the time-integrated Stokes parameters are obtained by integrating the time-resolved ones within the $T_{90}$.
\begin{equation}
\bar{F}=\frac{\int_{T_5}^{T_{95}}F dt_{obs}}{\int_{T_5}^{T_{95}}dt_{obs}}
\end{equation}
\begin{equation}
\bar{Q}=\frac{\int_{T_5}^{T_{95}}Q dt_{obs}}{\int_{T_5}^{T_{95}}dt_{obs}}
\end{equation}
\begin{equation}
\bar{U}=\frac{\int_{T_5}^{T_{95}}U dt_{obs}}{\int_{T_5}^{T_{95}}dt_{obs}}
\end{equation}
where $T_{5}$ and $T_{95}$ are the times when the time-accumulated flux reaches $5\%$ and $95\%$ of the total flux during the burst duration, respectively. And $T_{90}=T_{95}-T_{5}$. $t_{obs}$ is the observational time.

For an aligned field in the emission region, the time- and energy-integrated Stokes parameters $\bar{Q}$ and $\bar{U}$ are both nonzero, the final time- and energy-integrated PD ($\Pi$) and PA ($\chi$) of the GRB emission can be calculated as:
\begin{equation}\label{PI}
  \Pi=\frac{\sqrt{(\bar{Q}^2+\bar{U}^2)}}{\bar{F}},
\end{equation}
\begin{equation}\label{chi}
  \chi=\frac{1}{2}\arctan(\frac{\bar{U}}{\bar{Q}}).
\end{equation}
For a toroidal field, one of the Stokes parameters ($\bar{U}$) is zero, then the $\Pi$ is
\begin{equation}\label{PI}
  \Pi=\frac{\bar{Q}}{\bar{F}},
\end{equation}
and the $\chi$ for $\bar{Q}>0$ will have a $90^\circ$ difference with that for $\bar{Q}<0$.

It should be noted that the magnetic field in the emission region of the MR model would be very likely to be mixed because of the instability and/or the turbulent developed in the plasma. Since the profile of the PD curve with a mixed field will be very similar to that with the corresponding ordered field \citep{Lan2019,LD2020}, here we only assume that magnetic field is large-scale ordered in the radiation region. Hence, the time- and energy-integrated PD predictions ($\Pi$) in this paper would be the theoretical upper limits.

\section{Numerical results}
\label{Numerical results}

Because the current polarization analysis can only be done for bright bursts, in general the observation angle $\theta_V$, which is the angle between the line of sight and the jet axis, for the bursts with polarization detection should be within the jet cone, i.e., with $q\equiv\theta_V/\theta_j\leq1$. In statistics, the value of $\Pi$ at the plateau stage of the $q-\Pi$ curve with typical parameters when $q<1$ would be equal to the mean value of the time-integrated PD of the simulated bursts \citep{Toma2009,LWD2021}. So the predicted $q-\Pi$ curve with indicated typical parameters, especially for $q\leq1$, will be important. Although the flux density will drop dramatically when $q>1$ and it is impractically to measure the polarization at large off-axis angles, the $q-\Pi$ curves with $q>1$ are also studied for reference here.

With the model described in Section 2 and the Appendix, we numerically calculate the variations of the $\Pi$ with the parameter $q$. Without special illustration, parameters used in the following calculations will take the following fixed values: $\Gamma_0=100$, $\theta_j=0.1$ rad, $\alpha_s =-0.2$, $\beta_s=1.2$, $r_{on}=10^{14}$ cm, $r_{off}=3\times10^{16}$ cm, $r_0=10^{15}$ cm, $b=1$, and $B'_{0}=30$ G \citep{Ghirlanda2018,Lloyd2019,RE2023,Preece2000,Uhm_2018}. The orientation of the aligned field is set to be $\delta_a=\pi/6$. In Figures. 2, 4-5, the points correspond to our calculation results.

Single-energy electrons are assumed in the model here, as in the former studies \citep{Uhm_Zhang2015, Uhm_Zhang2016, Uhm_2018,LD2020}, their Lorentz factor is denoted as $\gamma_{ch}$. Two $\gamma_{ch}$ patterns are considered, corresponding to two general $E_p$ evolution mode (i.e., hard-to-soft and intensity-tracking modes). 

\begin{figure*}
  \centering
  \includegraphics[scale=0.6]{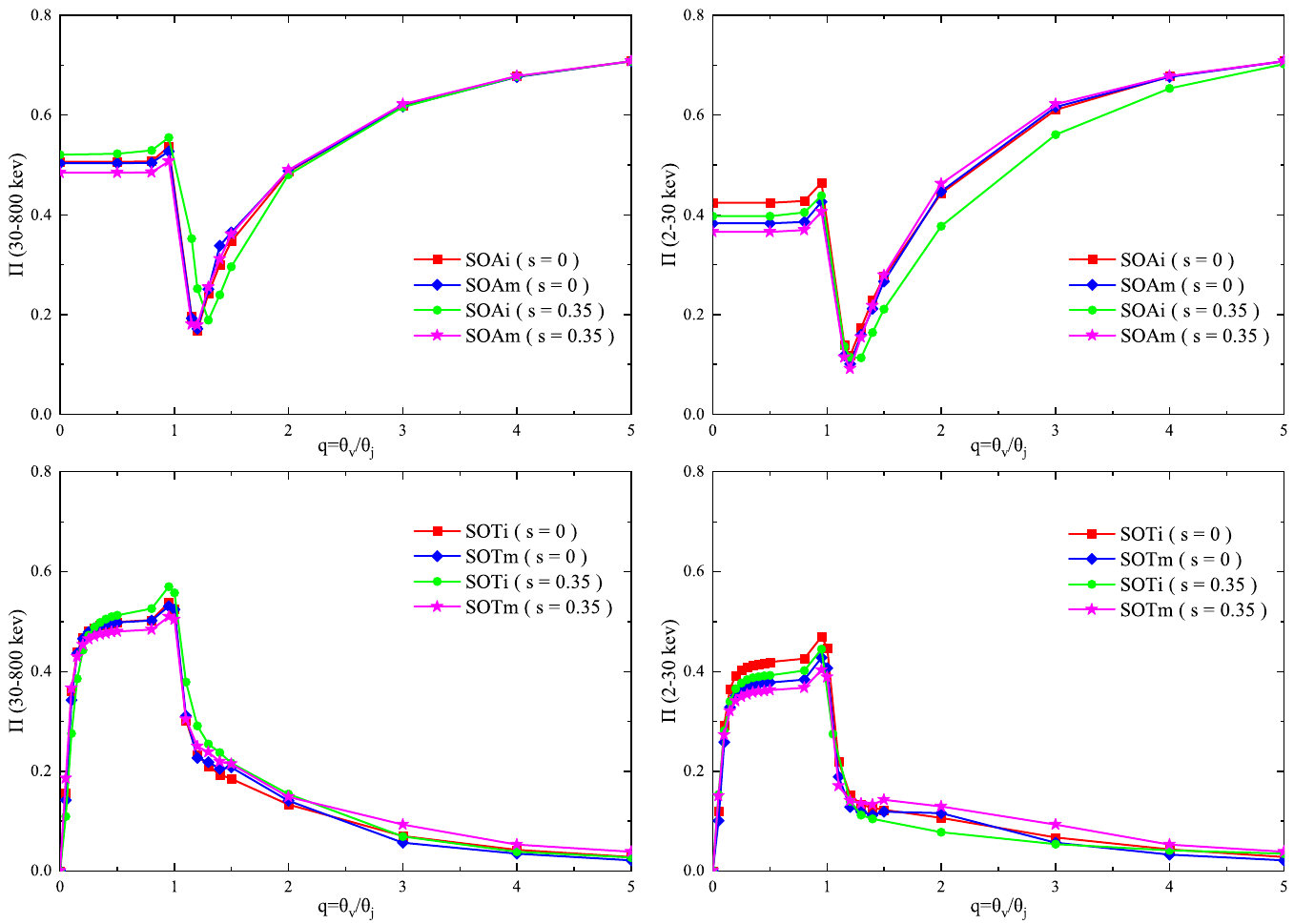}
  \caption{PD ($\Pi$) variations with $q\equiv\theta_V/\theta_j$. The left and right panels correspond to the results of the HPD (30-800 keV) and LPD (2-30 keV), respectively. The upper and lower panels are for the aligned and toroidal field in the emission region, respectively. In each upper panel, the red squares, green circles, blue diamonds, and pink stars represent the models of SOAi with $s=0$, SOAi with $s=0.35$, SOAm with $s=0$, and SOAm with $s=0.35$. In each lower panel, the red squares, green circles, blue diamonds, and pink stars represent the models of SOTi with $s=0$, SOTi with $s=0.35$, SOTm with $s=0$, and SOTm with $s=0.35$.}\label{Fig2}
\end{figure*}

First, we discuss the influences of the parameter $s$ and the $\gamma_{ch}$ patterns on the time- and energy-integrated PD ($\Pi$). The results are shown in Figure \ref{Fig2}. Total 8 models are compared, SOAi with $s=0$, SOAi with $s=0.35$, SOAm with $s=0$, SOAm with $s=0.35$, SOTi with $s=0$, SOTi with $s=0.35$, SOTm with $s=0$, and SOTm with $s=0.35$. Here, the "SO" represents the synchrotron emission in an ordered magnetic field, "A" or "T" are for the aligned or the toroidal field, and "i" or "m" correspond to the "i" (with a hard-to-soft $E_p$ evolution pattern) or "m" (with an intensity-tracking $E_p$ mode) mode.

There are two different variation profiles of the $q-\Pi$ curve for these 8 models in Figure \ref{Fig2}, corresponding to the two different ordered MFCs in the radiation region. The two profiles are independent of the bulk Lorentz factor pattern ($s$), the $E_p$ evolution mode, and the observational energy band. $\Pi$ values around $q=0$ will increase rapidly with $q$ for the toroidal field because the axial symmetry of the observed emission region was broken speedy, while for an aligned field there is no such symmetry and the asymmetry offered by the aligned field keeps roughly unchanged, hence $\Pi$ maintains a roughly constant value for $q<1$. And also because of different asymmetry, $\Pi$ will increase with $q$ to roughly 70\% when $q\gg1$ after a sudden decay for the aligned field, while it will roughly decay shallowly with $q$ to zero when $q\gg1$ after a sudden decay for the toroidal field.

To interpret the different variation trends of the aligned and toroidal fields at large $q$ value ($q\gg1$), we plot the local polarization distribution on the plane of sky for the SOAi and SOTi models with a constant velocity jet shell (i.e., $s=0$) at the peak time of the light curve of $q=3$. The observational energy is taken as 400 keV for reference \footnote{The local polarization distribution is similar for the observational energy of 15 keV.}. The other parameter are taken as their fixed values. For an aligned field, the local PD ($PD(\theta,\phi)$) and PA ($PA(\theta,\phi)$) of the emission from a point-like region \citep{Sari1999} are defined as follows.
\begin{equation}
  PD(\theta,\phi)=\frac{\sqrt{q_\nu(\theta,\phi)^2+u_\nu(\theta,\phi)^2)}}{f_\nu(\theta,\phi)},
\end{equation}
\begin{equation}
  PA(\theta,\phi)=\frac{1}{2}\arctan(\frac{u_\nu(\theta,\phi)}{q_\nu(\theta,\phi)}).
\end{equation}
where $f_\nu(\theta,\phi)$, $q_\nu(\theta,\phi)$ and $u_\nu(\theta,\phi)$ are the Stokes parameters of the point-like region. While for a toroidal field, one of the Stokes parameters ($u_\nu(\theta,\phi)$) is zero, then the local PD for a point-like region is
\begin{equation}
  PD(\theta,\phi)=\frac{q_\nu(\theta,\phi)}{f_\nu(\theta,\phi)},
\end{equation}
Its local PA will rotate abruptly by $90^\circ$ when the local PD changes its sign.
\begin{figure}
  \centering
  \includegraphics[scale=0.3]{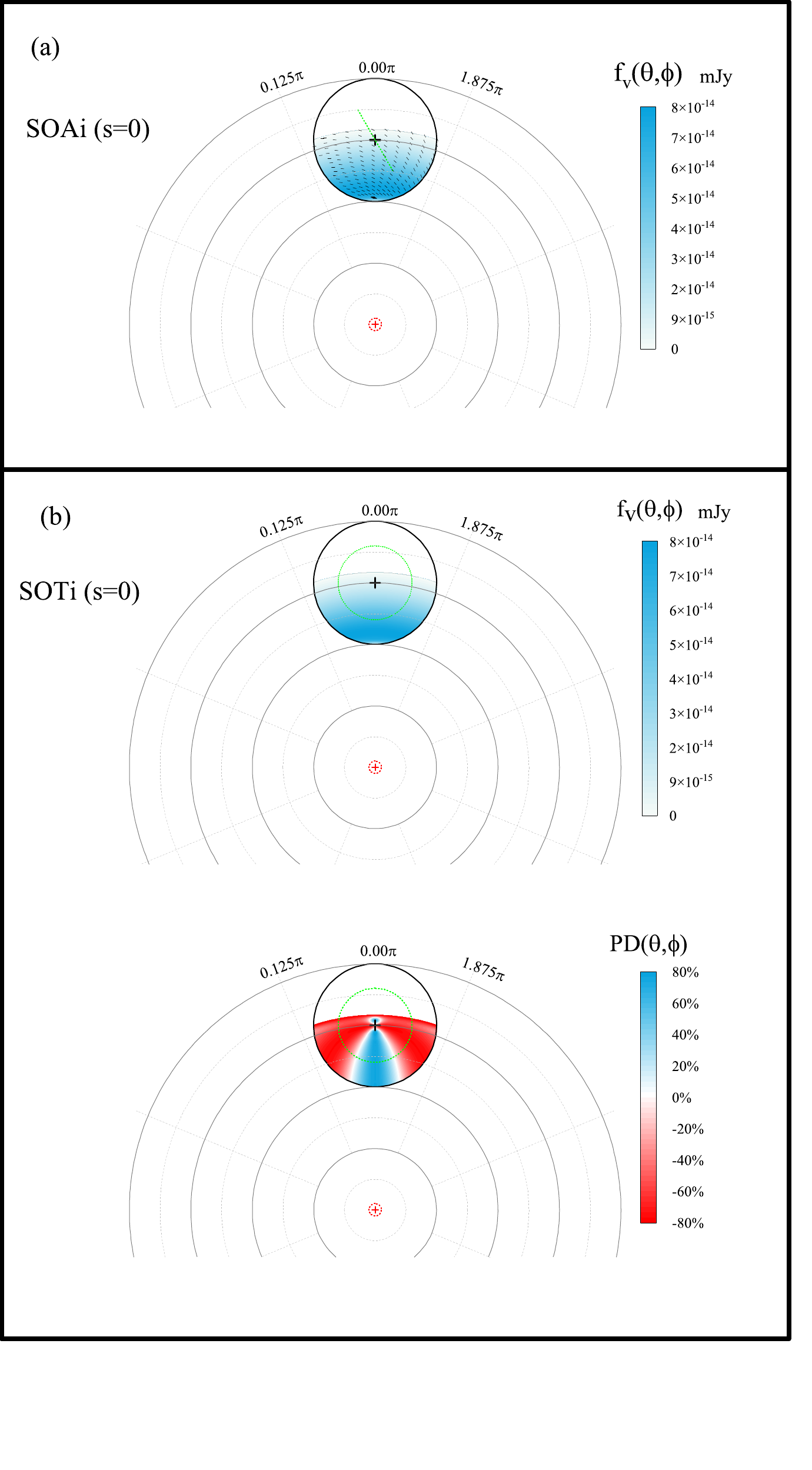}
  \caption{The distributions of the local flux ($f_\nu(\theta,\phi)$) and local polarizations on the plane of sky for SOAi model (a) and SOTi model (b). The red and black pluses show the projections of the LOS and jet axis. The red-dashed and thick-solid-black circles represent the projections of the $1/\Gamma_0$ cone and the jet cone. (a). The local PD of the emission from a point-like region is a constant of 0.77. The directions of the equivalent-length black short-lines show the directions of the local PA ($PA(\theta,\phi)$). The green dashed line represents the direction of the aligned field. (b). Upper panel shows the distribution of the local flux ($f_\nu(\theta,\phi)$). The lower panel shows the distribution of the local PD ($PD(\theta,\phi)$). The local PA ($PA(\theta,\phi)$) with a red color is rotated by $90^\circ$ compared with that of blue color. The green dashed circle shows the field lines of the toroidal field.}
\label{Fig3}
\end{figure}

For the aligned field, its local PD ($PD(\theta,\phi)$) is a constant and equals to $(\beta_s+1)/(\beta_s+5/3)=0.77$ for the observational energy of 400 keV. Since the syntropy of the local PA ($PA(\theta,\phi)$) is good at the regions where the local flux ($f_\nu(\theta,\phi)$) remains a relatively high value, the PD of the jet radiation ($\Pi$) with an aligned field would be as large as $60\%$ at $q=3$ for HPD. While for the toroidal field, the local flux also decrease with $\theta$ and the polarized flux are almost been cancelled. So the PD of the jet radiation with a toroidal field would be small.

There are only small differences in concrete $\Pi$ value in each panel, and the influences of the bulk Lorentz factor pattern ($s$) and the $E_p$ evolution mode on the time- and energy-integrated PD are very limited. Therefore, in the following we will focus on the "i" model with $s=0$ to discuss the influences of other parameters on $\Pi$. It should be noted that $\Pi$ will be larger for HPD than LPD at the plateau of the $q-\Pi$ curve when $q<1$. Since in statistics the value of $\Pi$ at the plateau stage of the $q-\Pi$ curve with typical parameters when $q<1$ would be equal to the observed mean value of the time-integrated PD \citep{LWD2021}, the detected mean value of HPD will be larger than that of LPD. This result is consistent with that in \cite{GL2023}.

\begin{figure*}
  \centering
  \includegraphics[scale=0.6]{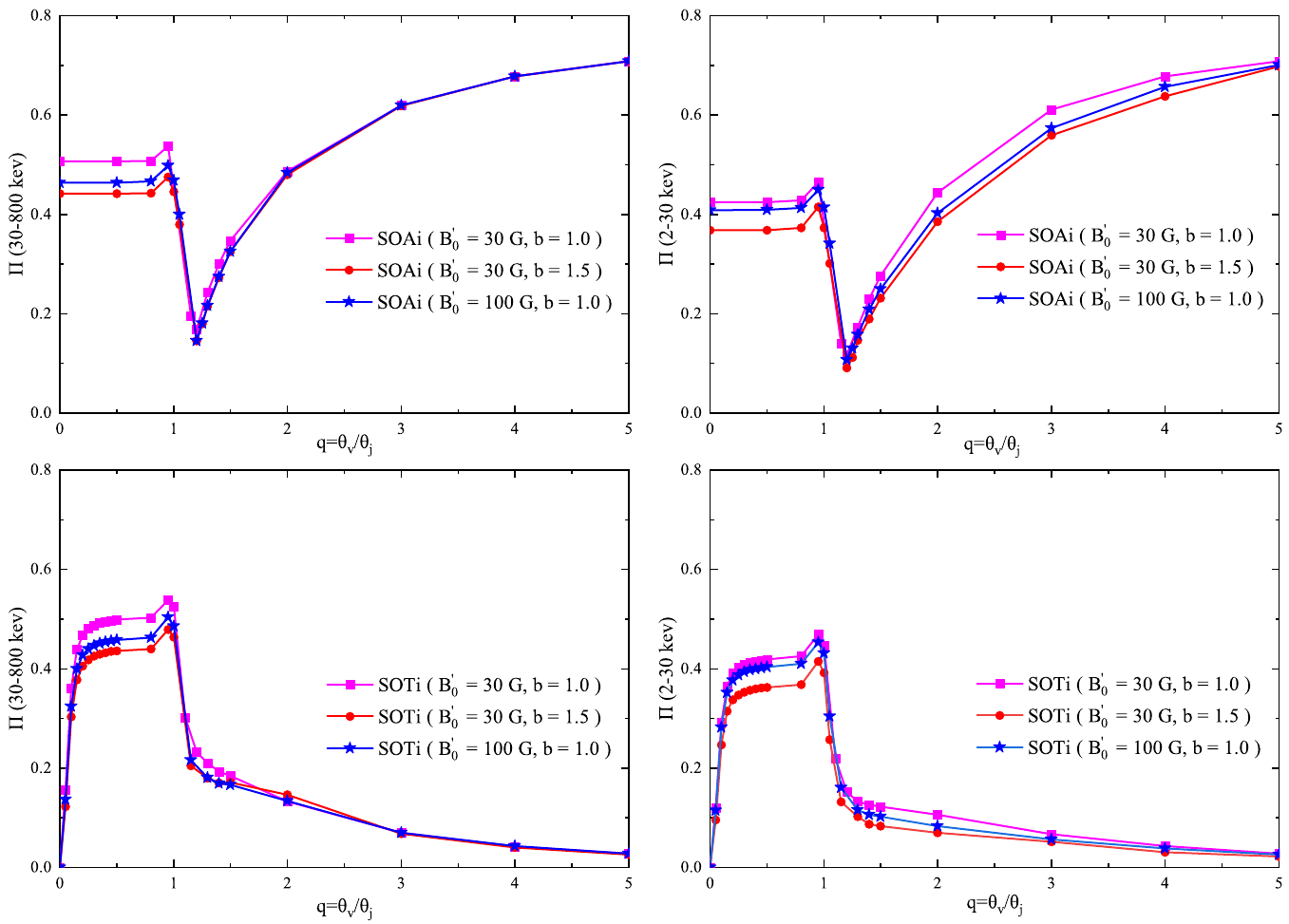}
  \caption{Same as Figure 2, but only the models with "i" mode and $s=0$ are considered. In each panel, the pink squares, the red circles, and blue stars correspond to the ($B'_0$, b) sets of (30 G, 1), (30 G, 1.5), (100 G, 1).}
\label{Fig4}
\end{figure*}

Then the influences of the magnetic field strength on the time- and energy-integrated PD are investigated. The results are shown in Figure \ref{Fig4}. Three parameter sets of ($B'_0$, $b$) are considered \citep{Uhm_Z.Lucas2014}, i.e., (30 G, 1), (30 G, 1.5), (100 G, 1). Again, there are two $q-\Pi$ profiles corresponding to two MFCs. And different values of both $B'_0$ and decay index $b$ will not affect the $q-\Pi$ profile and have limited influences on the $\Pi$ value. We note that, for example for the HPD, there are $\leq10\%$ difference for the time-integrated PDs at the plateau phase when $q<1$ for the different combinations of ($B'_0$, $b$). Because the time-integrated flux and polarized flux are mainly from the peak time of the light curve and the PA is roughly a constant for on-axis observations, time-integrated PD of the single-pulse burst is roughly equal to the time-resolved PD at the time of the light-curve peak. So we use the PD at peak time to investigate. We found that the influence of different combinations of ($B'_0$, $b$) on the distribution of the local flux density on the EATS will be larger than that for the local polarized flux \footnote{The local flux density and the local polarized flux are defined as the flux density and the polarized flux from one $\theta-$circle at the EATS here.}, which will result in a different time-resolved PD at the peak time for different combinations of ($B'_0$, $b$), so does the time-integrated one. And we will simply take $B'_0=30$ G and $b=1$ in our following analysis.

\begin{figure*}
  \centering
  \includegraphics[scale=0.6]{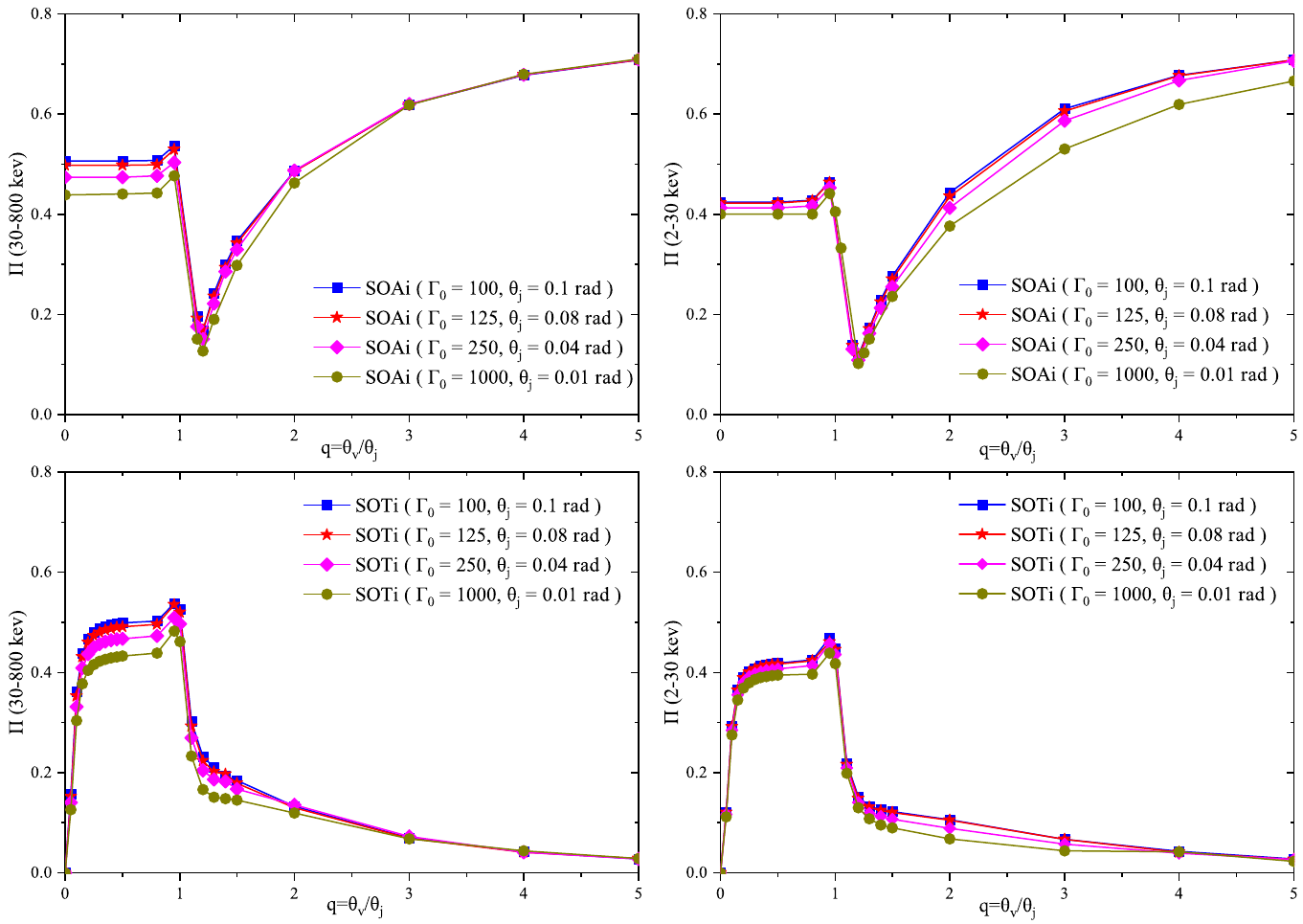}
  \caption{Same as Figure 2, but the product of $\Gamma_0\theta_j$ is a constant value of 10 and only the models with "i" mode and $s=0$ are considered. In each panel, the blue squares, red stars, pink diamonds, and brown circles are for the ($\Gamma_0, \theta_j$) sets of (100, 0.1 rad), (125, 0.08 rad), (250, 0.04 rad), and (1000, 0.01 rad).}
\label{Fig5}
\end{figure*}

In Figure \ref{Fig5}, the influences of different sets of ($\Gamma_0, \theta_j$) with a constant $\Gamma_0\theta_j$ value are studied. Since the estimated values of $\theta_j$ and $\Gamma_0$ are around 0.1 rad and 100 \citep{Lloyd2019,RE2023,Ghirlanda2018}, here we take $\Gamma_0\theta_j=10$. As our above studies, depending on the two MFCs, there are two $q-\Pi$ profiles and different set of ($\Gamma_0, \theta_j$) with a $\Gamma_0\theta_j$ value of 10 will not affect the $q-\Pi$ profile and has a limited influence on $\Pi$ value. Same as the case with different combinations of ($B'_0$, $b$), there are also  $\leq10\%$ difference for the time-integrated PDs at the plateau phase when $q<1$ for the different sets of ($\Gamma_0, \theta_j$). With the study, same as that for the combinations of ($B'_0$, $b$), the influence of different sets of ($\Gamma_0, \theta_j$) on the distribution of the local flux density on the EATS will be larger than that for the local polarized flux, which will result in a different time-resolved PD at the peak time for different sets of ($\Gamma_0, \theta_j$), so does the time-integrated one. The variations of the $q-\Pi$ curve with both $\Gamma_0$ and $\theta_j$ are also investigated, respectively. Independent to the concrete values of $\Gamma_0$ or $\theta_j$, the profiles of the $q-\Pi$ curves are similar if the values of $\Gamma_0\theta_j$ are the same.

In statistics, the predicted PDs with a large-scale ordered magnetic field in its radiation region will concentrate around $(40-50)\%$ for HPD and around $(30-40)\%$ for LPD, while it is around $(20-40)\%$ for HPD and around $(15-30)\%$ for LPD in the former estimations \citep{GL2023}. The predicted PD here is higher than that of the former estimation. That is because the variations of magnetic field strength, bulk Lorentz factor, and the Lorentz factor of electrons in the comoving frame with radius would also affect the final time- and energy-integrated PD, as shown in Figures. 2, 4-5.

\section{Conclusion and Discussion}\label{sec:discussion}

The time evolution information was missed in the former studies of the time-integrated PD in GRB prompt phase \citep{Granot2003b,Lyutikov2003,Nakar2003,GT2005,Toma2009,GL2023}. Here, it is considered with the time-resolved polarization model proposed in \cite{LD2020}. The model can mimic both the emission from the MR model with an accelerating or a constant-velocity shell and IS model with a constant-velocity shell. Here, in addition to a toroidal field, a large-scale aligned field is also considered, which is also one of the popular MFC in the literature \citep{Spruit2001,Drenkhahn2002}. The profile of the $q-\Pi$ curve for an aligned field is very different from that of a toroidal field. The results here are consistent with that in \cite{Toma2009} and \cite{GL2023}. 

The statistical properties of the time-integrated PD can be read roughly from the $q-\Pi$ curve \citep{Toma2009,LWD2021}. The time-integrated PDs of the most simulated bursts will concentrate on the $\Pi$ value at the plateau stage of the $q-\Pi$ curve with typical values of the parameters when $q<1$ \citep{LWD2021}. Most of the GRBs used for polarization analyse are very bright, indicating on-axis observations. The observed time-integrated PDs should also concentrate on the PD values at the plateau of the $q-\Pi$ curve with typical values of the parameters for $q<1$. It is known that the local PD in an ordered field will be larger for high-energy spectral index $\beta$ than low-energy spectral index $\alpha$ \citep{RL1979}. Since the high-energy photons with large spectral index (i.e., $\beta$) become less when $E_p$ increases, $\Pi$ of the jet emission will decrease with $E_p$ \citep{Toma2009,GL2023}. Hence, the statistical properties do not need to be studied specifically.

Here, the predicted PAs are roughly constant within $T_{90}$ for on-axis observations with $q\leq1$ and the magnetic field is assumed to be large-scale ordered. Therefore, the predicted time-integrated PD of GRB prompt phase is the upper limit, which is around $(40-50)\%$ with typical parameters for HPD and is larger than that of $(20-40)\%$ the former studies \citep{Toma2009,GL2023}. Hence, the upper limit of $\Pi$ calculated here could also interpret most of the current time-integrated PDs observed by GAP, POLAR, and AstroSat \citep{Yonetoku2011,Yonetoku2012,Zhang2019,Kole2020,Sharma2019,Chattopadhyay2022}. The predicted upper limit of $\Pi$ at 250 keV is around 25\% in \cite{LWD2021} (see Fig. 1 in their paper), that is because their derivation of the comoving peak frequency ($h\nu'_0=E_p(1+z)/\mathcal{D}$) is wrong. However, the profiles of their $q-\Pi$ curve for the aligned and toroidal fields are similar to our results here. The $\Pi$ value here for the LPD at the plateau stage of the $q-\Pi$ curve will be relatively smaller ($(30-40)\%$) than that for the HPD, and the result for LPD here is also larger than that of $(15-30)\%$ in \cite{GL2023}.

The profile of the $q-\Pi$ curve depends mainly on both the MFC in the emitting region. Because the asymmetry offered by the large-scale ordered aligned magnetic field is roughly unchanged when $q\leq1$, $\Pi$ values around $q=0$ are also constants as in the $\Pi$ plateau phase when $0<q<1$ for an aligned field, while it is very different for a toroidal field with a sharp rise of $\Pi$ starting from 0 when $q\geq0$ followed by a plateau phase when $q<1$. Also due to different asymmetries, roughly when $q>1+1/(\Gamma_0\theta_j)$, $\Pi$ will increase with $q$ to about 70\% for the aligned field, while it decays shallowly to zero for the toroidal field. Although the comoving peak frequency is incorrect in \cite{LWD2021}, the profiles of their $q-\Pi$ curve for the aligned and toroidal field are similar to the results here. 

Although the calculation methods are different, our profile of the $q-\Pi$ curve for the toroidal field is consistent with that in \cite{Toma2009}. Although both the time-integrated flux density and the time-integrated $E_p$ are mainly determined by emission episode around the light-curve peak, and PA for on-axis observations is a constant within $T_{90}$ \citep{WL2023}, the variations of the parameters on the EATS would also affect the value of $\Pi$. So the predicted $\Pi$ here is larger than that estimated via the former time-integrated method. Therefore, the calculation method of the time-integrated PD proposed in \cite{Granot2003b,Lyutikov2003,Nakar2003,GT2005,Toma2009} and \cite{GL2023} can only be a preliminary estimation. Although the predicted PAs are almost constants during the burst duration for the on-axis observations, it will rotate for slightly off-axis observations \citep{WL2023}, for which the flux is roughly comparable to the on-axis one. Hence, the time-resolved method has the potential to interpret the observations with PA rotation \citep{WL2023} and will be investigated in detail in our future work.

\section*{Acknowledgements}
This work is supported by the National Natural Science Foundation of China (grant Nos. 11903014) , M.X.L also would like to appreciate the financial support from Jilin University.

\section*{Data Availability}
The data underlying this work will be shared on reasonable request to the corresponding author.



\bibliographystyle{mnras}
\bibliography{ms_arXiv} 


\appendix

\section{The formula of the time- and energy-resolved Stokes parameters}

The corresponding expressions of the time- and energy-resolved Stokes parameters ($F_{\nu}$, $Q_{\nu}$, and $U_{\nu}$) of the jet radiation are shown as follows \citep{LD2020}.
\begin{equation}\label{F_v}
F_{\nu}=\frac{1+z}{4 \pi D_{L}^{2}} \int \mathcal{D}^{3} \sin \theta d \theta \int  \frac{N P_{0}^{\prime} H_{e n}(x)\sin \theta'_{B}}{4 \pi}d \phi,
\end{equation}
\begin{equation}\label{Q_v}
\begin{split}
Q_{\nu}= & \frac{1+z}{4 \pi D_{L}^{2}} \int \mathcal{D}^{3} \sin \theta d \theta \\
& \times \int \Pi_{p,b} \cos 2 \chi_{p} \frac{N P_{0}^{\prime} H_{e n}(x)\sin \theta'_{B}}{4 \pi}d \phi ,
\end{split}
\end{equation}
\begin{equation}\label{U_v}
\begin{split}
U_{\nu}= & \frac{1+z}{4 \pi D_{L}^{2}} \int \mathcal{D}^{3} \sin \theta d \theta \\
& \times \int  \Pi_{p,b} \sin 2 \chi_{p} \frac{N P_{0}^{\prime} H_{e n}(x)\sin \theta'_{B}}{4 \pi}d \phi.
\end{split}
\end{equation}
where $z$ is the redshift, $D_L$ is the luminosity distance of the source, and the Doppler factor is $\mathcal{D}=1/\Gamma(1-\beta\cos\theta)$. $\Gamma$ is the bulk Lorentz factor. The isotropic total electron number in the shell is $N=\int{R_{inj} dt}/{\Gamma}$. Electrons are assumed to be fed into the shell at an isotropic rate $R_{inj}$. $\phi$ represents the angle in the plane of the sky between the projection of the jet axis and that of a local fluid element's radial direction.

$P'_0 $ represents the magnitude of the spectral power of a single electron and $H_{en}(x)$ describes the shape of the emission spectrum.
\begin{equation}
P'_0=\frac{3\sqrt{3}}{32}\frac{m_ec^2\sigma_TB'}{q_e}
\end{equation}
where $m_e$ and $q_e$ are the mass and charge of the electron, $c$ is the speed of the light, $\sigma_T$ represent the Thomson cross section, and $B'$ is the magnetic field strength in the comoving frame.
\begin{equation}
H_{en}(x)=\begin{cases}
x^{-\alpha}exp(-x), &\text{$x\leq x_c$}, \\ 
x_c^{x_c}exp(-x_c)x^{-\beta}, &\text{$x\geq x_c$},
\end{cases}
\end{equation}
where $x_c=\beta-\alpha$ and $x=\nu'/\nu'_{ch}$. And $\alpha$ and $\beta$ are the low- and high-energy spectral index of the Band function. $\nu'$ and $\nu'_{ch}$ are the observed and critical frequencies in the comoving frame. And $\nu'_{ch}=q_eB'\gamma^2_{ch}\sin \theta'_{B}/2\pi m_ec$. 

Two kinds of $\gamma_{ch}$ patterns are considered here, namely the "i" and "m" models. For the "i" model, the pattern is a single power of radius.
\begin{equation}
\gamma_{ch}(r)=\gamma_{ch}^0(r/r_0)^g,
\end{equation}
where we take $\gamma_{ch}^0=5\times10^4$ and $g=-0.2$. For the "m" model, the profile of the Lorentz factor of the electrons is a broken power law with respect to the radius.
\begin{equation}
\gamma_{ch}(r)=\gamma_{ch}^m\times\begin{cases}
(r/r_m)^g, & \text{$r\leq r_m$}, \\ (r/r_m)^{-g}, & \text{$r\geq r_m$},
\end{cases}
\end{equation}
where we take $r_m=2\times10^{15}$ cm and $\gamma_{ch}^m=2\times10^5$, and the power-law index is taken as $g=1.0$.

A broken local PD in an ordered magnetic field $\Pi_{p,b}$ is used \citep{Toma2009,WL2023}.
\begin{equation}\label{equa}
\Pi_{p,b}=\frac{\tilde{\alpha}+1}{\tilde{\alpha}+5/3}
\end{equation}
where $\tilde{\alpha}$ equals to the spectral index of the Band spectrum with $\tilde{\alpha}=\alpha_s$ for low-energy spectrum and $\tilde{\alpha}=\beta_s$ for high-energy spectrum.

For the specific form of $\sin \theta'_B$ and the local PA $\chi_{p}$ for an aligned field can be found in Eqs. (25) and (26) in \cite{LWD2016} and for a toroidal field can be found in Eqs. (22) and (23) also in \cite{LWD2016}. The integrations for the time-resolved Stokes parameters are done on the equal arrival time surface (EATS). The shell starts to emit photons at radius $r_{on}$ (corresponding to the burst source time $t_{on}$). The photons emitted from radius r at burst source time t with a polar angle $\theta$ with respect to the observer's line of sight will be received by an observer at time $t_{obs}$.
\begin{equation}\label{tobs}
t_{obs}=\begin{cases}
[t-\frac{r}{c}\cos\theta-t_{on}+\frac{r_{on}}{c}](1+z), &\text{$\theta_V\leq \theta_j$}, \\ [t-\frac{r}{c}\cos\theta-t_{on}+\frac{r_{on}}{c}\cos(\theta_V-\theta_j)](1+z), &\text{$\theta_V\geq \theta_j$}.
\end{cases}
\end{equation} 
where $\theta$ is the angle between the velocity of the jet element and the line of sight in the observer frame.


\bsp	
\label{lastpage}
\end{document}